\documentclass[12pt]{article}
\usepackage{epsf} 
\usepackage{amsmath}
\usepackage{amssymb}
\usepackage{epsfig}
\usepackage{latexsym}
\usepackage{amsfonts}
\usepackage{graphicx}%
\usepackage{varioref}
\usepackage{ifthen}
\begin{document}
\parindent 0mm 
\setlength{\parskip}{\baselineskip} 
\thispagestyle{empty}
\pagenumbering{arabic} 
\setcounter{page}{0}
\mbox{ }
\rightline{UCT-TP-293/13, MITP/13-026}
\newline
\rightline{April 2013}
\newline
\vspace{0.1cm}
\begin{center}
{\large {\bf 	Strange quark mass from sum rules with improved perturbative QCD convergence}}
{\large \footnote{{\large {\footnotesize Supported in part by  NRF (South Africa) and Alexander von Humboldt Foundation (Germany).}}}}\\
\end{center}
\vspace{.05cm}
\begin{center}
{\bf Sebastian Bodenstein}$^{(a)}$, \,
{\bf  Cesareo \;  A. \; Dominguez}$^{(a)}$  {\bf and} \\
{\bf Karl Schilcher}$^{(a),(b)}$
\end{center}

\begin{center}
{\it $^{(a)}$Centre for Theoretical Physics and Astrophysics\\[0pt]University of
Cape Town, Rondebosch 7700, South Africa\\
$^{(b)}$ Institut f\"{u}r Physik, Johannes Gutenberg-Universit\"{a}t\\
Staudingerweg 7, D-55099 Mainz, Germany}
\end{center}

\vspace{0.2cm}
\begin {footnotesize}
{\it E-mail:} sebastianbod@gmail.com, cesareo.dominguez@uct.ac.za, karl.schilcher@uni-mainz.de
\end{footnotesize}

\begin{center}

\textbf{Abstract}
\end{center}
\noindent
The strange quark mass is determined from a  QCD Finite Energy Sum Rule (FESR) optimized to reduce considerably the systematic uncertainties arising from the  hadronic resonance sector, as well as from the poor convergence of  the pseudoscalar correlator in perturbative QCD. The former is achieved by introducing a suitable integration kernel in the Cauchy integral in the complex squared energy plane. The latter is obtained by optimizing the perturbative expansion to accelerate its convergence. The result for the strange quark mass in the $\overline{MS}$ scheme at a scale of $2 \;{\mbox{GeV}}$ is
$\overline{m}_s (\mbox{2 GeV}) = 94 \,\pm \, 9\; \mbox{MeV}$.

\newpage
\bigskip
\noindent
\section{Introduction}
\noindent

The framework of QCD sum rules \cite{Review} allows for an analytic determination of light-quark masses by relating them e.g. to  light pseudoscalar meson poles and their radial excitations   using the pseudoscalar correlator in QCD (for recent reviews see \cite{PDG}-\cite{CAD_R}). The hadronic sector is intimately related to QCD and the operator product expansion (OPE) by means of Cauchy's theorem in the complex squared energy plane, i.e. quark-hadron duality. However, unlike the case of heavy-quarks where the hadronic (vector) correlator is directly known from data, the pseudoscalar two-point function is not experimentally known beyond the pole, except for the masses and widths of the first two radial excitations of the pion and the kaon. This information is hardly enough to fully reconstruct the hadronic spectral function. Indeed, one first needs to invoke a specific model parametrization of the resonances, e.g. a Breit-Wigner form, for which there is no direct evidence. Next, potential inelasticity and non-resonant background are realistically impossible to model. As a result, (unknown) systematic uncertainties have affected light-quark mass determinations from QCD sum rules over the years. A major progress in reducing considerably this systematic uncertainty has been made recently by introducing integration kernels in the sum rules designed to strongly quench the contribution from the hadronic resonance sector \cite{DNRS1}. The result in this framework \cite{DNRS1} for the strange-quark mass in the $\overline{MS}$ scheme at a scale $\mu= 2\;{\mbox{GeV}}$, at five-loop order in PQCD is 
\begin{equation}
\overline{m}_s (2 \;\mbox{GeV}) = 102 \pm 8 \;{\mbox{MeV}}\;,
\end{equation}
in good agreement with contemporary \footnote{For earlier determinations based on three- or four-loop order PQCD, and/or superseded values of $\alpha_s$, see \cite{PDG}.} independent determinations \cite{Other}-\cite{CHET_K}. The main source of uncertainty in this result arises from the strong coupling (the value $\alpha_s(M_\tau^2) = 0.344 \pm 0.009$ from \cite{Davier1} was used in \cite{DNRS1}), followed by the hadronic resonance model which contributes less than $3\%$ using quenching integration kernels.\\ 
With the hadronic sector under better control, one needs to address another potential systematic uncertainty arising from the well known poor convergence of the pseudoscalar correlator in perturbative QCD (PQCD). The pseudoscalar correlator is defined as
\begin{equation}
\psi_{5} (q^{2})   = i \, \int\; d^{4}  x \; e^{i q x} \; 
<|T(\partial^\mu A_{\mu}(x) \;, \; \partial^\nu A_{\nu}^{\dagger}(0))|> \;,
\end{equation}
where $\partial^\mu A_{\mu}(x) = (m_s + m_{ud}) :\overline{s}(x) \,i \, \gamma_{5}\, u(x):\;$ is the divergence of the  axial-vector current, and $m_{ud} \equiv (m_u + m_d)/2$. The second derivative of $\psi_5(q^2)$ at a scale $\mu^2 = Q^2 \equiv -q^2$  in the $\overline{MS}$ scheme, to five-loop order in PQCD \cite{CHET_K} is given by
\begin{equation}
\psi_{5}^{"} (Q^{2})^{PQCD} = \frac{3}{8 \, \pi^2} \frac{(\overline{m}_s + \overline{m}_{ud})^2}{Q^2}
\left[1 +  3.7\, a + 14.2\, a^2 + 77.4\, a^3 + 512.0\, a^4\right] \;,
\end{equation}
where $a \equiv \alpha_s(Q^2)/\pi$. This behaviour is already providing a strong hint of a potential systematic uncertainty. If present, it could change both the central value as well as the error in 
the result for the strange quark mass, Eq.(1).
In \cite{DNRS1}  the unknown six-loop order contribution  was assumed to be as large as the five-loop one, resulting in a $3 \%$ uncertainty, included in Eq.(1). However, this can only be a guess. After trying a variety of well established methods for the acceleration of the convergence of series, none gave a satisfactory improvement for the pseudoscalar correlator. It must be kept in mind, though, that what actually matters is the convergence of the expression for the quark mass. This is the subject of this paper, where we describe a simple but successful approach to obtain a reasonably convergent series expansion for the strange quark mass in the framework of Fixed Order Perturbation Theory (FOPT). 
\section{Finite Energy Sum Rules}
Invoking Cauchy's theorem in the complex squared energy s-plane one obtains the Finite Energy Sum Rule (FESR)
\begin{eqnarray}
0 &=&
\int_{0}^{s_0}
ds \; p_i(s) \; \frac{1}{\pi} Im \;\psi_{5}(s) + \frac{1}{2\pi i}
\oint_{C(|s_0|)}
ds \; p_i(s) \;\psi_{5}(s) \nonumber \\ [.3cm]
&\simeq&
\int_{0}^{s_0}
ds \; p_i(s) \; \frac{1}{\pi}\, Im \;\psi_{5}^{HAD}(s) + \frac{1}{2\pi i}
\oint_{C(|s_0|)}
ds \; p_i(s) \;\psi_{5}^{QCD}(s) \, ,\label{CAUCHY1}
\end{eqnarray}
where $p_i(s)$ are analytic kernels, and  the contour integral is performed over a circle of radius $|s_0|$ large enough so that  $\psi_{5}(s)$ can  be safely replaced by its QCD counterpart $\psi_{5}^{QCD}(s)$. The first requirement in choosing an integration kernel is that is should quench the contribution of the two radial excitations of the kaon, i.e. the $K_1(1460)$ and the $K_2(1830)$, both with widths $\Gamma \simeq 250 \,{\mbox{MeV}}$. The simple choice 
\begin{equation}\label{EQ:p1}
p_1(s) = (s - M_{K_1}^2) (s - M_{K_2}^2) \;,
\end{equation}
was shown in \cite{DNRS1} to substantially quench this badly known contribution. A second requirement arises from potential duality violations which call for pinched kernels \cite{PINCH}. Hence, we will also consider
\begin{equation}\label{EQ:p2}
p_2(s) = (s - M_{K_1}^2) (s - M_{K_2}^2) (s_0 - s)\;.
\end{equation}
In \cite{DNRS1} this type of kernel was found to affect somewhat the stability of results for the strange quark mass against changes in  the PQCD threshold $s_0$. However, this is not the case with the method  to be introduced here, as will be discussed later. We do not consider kernels involving higher powers of $s$, as they would involve higher dimensional vacuum condensates in the OPE.  A third kernel to be  considered is one that vanishes at $s_0$, as well as somewhere between the $K_1(1460)$ and the $K_2(1830)$ resonances, i.e.
\begin{equation}\label{EQ:p3}
p_3(s) = (s - a) (s -s_0) \;,
\end{equation}
where the free parameter $a$ can be determined by demanding  maximally reduction of the uncertainty.
This kernel reduces the dependence of the quark mass on  $s_0$, it suppresses the contribution of the two resonances, and it does not call for higher-dimensional condensates ($d > 6$). \\
Concerning the modeling of the hadronic spectral function, the first requirement is that it should satisfy the threshold behaviour given by chiral perturbation theory (CPT), as first proposed long ago in \cite{CAD0}. This constraint also serves to fix the overall normalization of the resonance parametrization. For the case of the pseudoscalar correlator, Eq.(2), this threshold behaviour is given by \footnote{There is a misprint in the expression given in Eq.(18) of \cite{KPIPI}.} \cite{KPIPI}
\begin{equation}
\frac{1}{\pi} \, {\mbox{Im}} \, \psi_5(s)|_{K \pi \pi} = \frac{M_K^4}{2 f_\pi^2}\; \frac{3}{2^8  \pi^4} \; \frac{I(s)}{s (M_k^2 -s)^2} \, \theta (s - M_K^2) \;,
\end{equation}
where $I(s)$ is a known integral function \cite{KPIPI} accounting for the $K^*(892) - \pi$ sub-channel, of numerical importance given the narrow width of the $K^*(892)$ (other resonance sub-channels are negligible). In this expression the chiral $SU(2) \times SU(2)$ limit ($M_\pi =0$) was assumed. The hadronic spectral function can then be written as
\begin{eqnarray}
\frac{1}{\pi} {\mbox{Im}} \, \psi_5^{HAD}(s)&=&2 f_K^2 M_K^4 \delta(s - M_K^2) + \frac{1}{\pi}  {\mbox{Im}} \, \psi_5(s)|_{K \pi \pi} \frac{[BW_1(s) + \lambda \,BW_2(s)]}{(1 + \lambda)}
\nonumber \\[.3cm]
&+& \frac{1}{\pi} \, {\mbox{Im}} \, \psi_5(s)|_{\mbox{\tiny{QCD}}} \;\theta(s - s_0) \;,
\end{eqnarray}
where $f_K = 110.4 \pm 6 \, {\mbox{MeV}}$ from \cite{PDG}, and the parameter $\lambda$, of order unity, controls the relative weight of the two resonances. 
We shall also consider an alternative resonance parametrization \cite{KM} which does not have the CPT constraint. To be conservative we consider a $33\%$ overall normalization uncertainty in the model.\\

In the QCD sector the pseudoscalar correlator is known in PQCD to five-loop order, and the Wilson coefficients in the OPE are known for the dimension $d=2$ quark mass correction, the gluon and the quark condensates, and the quartic quark mass corrections \cite{CHET_K}. Information on the  the gluon condensate, and the higher dimensional condensates has been obtained from analyses of $\tau$-decay \cite{tau} and $e^- e^+$ annihilation into hadrons \cite{e+e-}. The quark condensates have been determined from the Gell-Mann-Oakes-Renner relation,  including its corrections \cite{GMOR}.
The FESR, Eq.(4), in the framework of Fixed Order Perturbation Theory (FOPT) can then be written as
\begin{equation}
\delta_5^{HAD}(s_0) = \left[\bar{m}_s(s_0) + \bar{m}_{ud}(s_0)\right]^2 \; \delta_5^{QCD}(s_0) \;,
\end{equation}
where
\begin{equation}
\delta_5^{HAD}(s_0) = \int_{0}^{s_0}
ds \; p_i(s) \; \frac{1}{\pi}\, Im \;\psi_{5}^{HAD}(s) \;,
\end{equation}
and $\delta_5^{QCD}(s_0)$ is defined here as \footnote{Notice that this definition is different from that in \cite{DNRS1}.}
\begin{equation}
\delta_5^{QCD}(s_0) = - \frac{1}{2\pi i} \oint_{C(|s_0|)} ds \; p_i(s) \;\psi_{5}^{OPE}(s) \;,
\end{equation}
where for later convenience  the quark masses have been factored out of $\psi_5^{QCD}(s)$, and thus $\psi_{5}^{OPE}(s)$ is  the remainder in PQCD plus power corrections in the OPE. The quark mass is then determined from the FESR
\begin{equation}
\bar{m}_s(s_0) = \left[1 + \frac{\bar{m}_{ud}}{\bar{m}_s}\right]^{-1}\left[\frac{\delta_5^{HAD}(s_0) }{\delta_5^{QCD}(s_0)}\right]^{1/2}\;,
\end{equation}
where the quark mass ratio $\bar{m}_{ud}/\bar{m}_s$ is an input from e.g. CPT or Lattice QCD (LQCD) \cite{FLAG}.\\

\section{Results}

The hadronic spectral function, Eq.(9), is plotted in Fig. 1 (curve (a)), together with the model of \cite{KM} (curve(b)), and the PQCD expression of $\psi_5(s)$ to five-loop order (curve (c)) using \cite{PDG}
\begin{equation}
\alpha_s(M_Z) = 0.1184 \pm 0.0007 \;,
\end{equation}
scaled down to $\alpha_s(s_0)$. This value is  in perfect agreement with a recent world average \cite{PICH} $\alpha_s(M_Z) = 0.1186 \pm 0.0007$. \\ 
\begin{figure}
[ht]
\begin{center}
\includegraphics[height=2.5in, width=4.5in]
{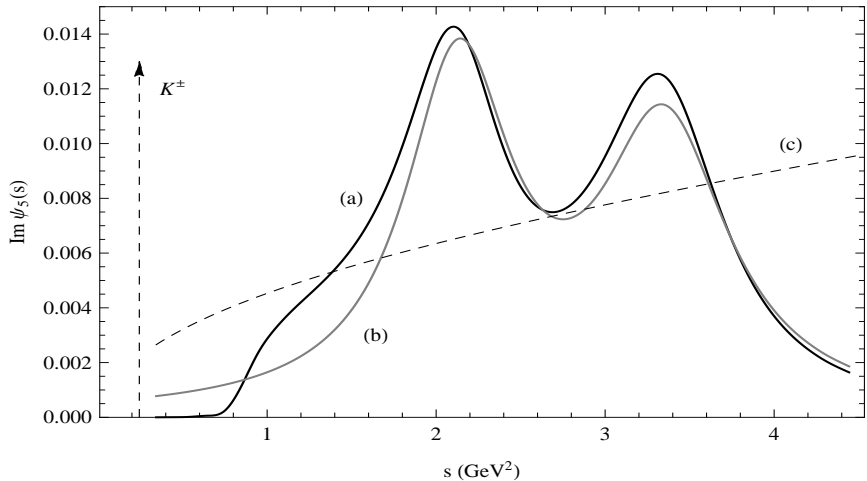}
\caption{\footnotesize{The hadronic spectral function, Eq.(9) with $\lambda = 1$, curve (a), and the model of \cite{KM}, curve (b). Curve (c) is the PQCD result for $\psi_5(s)$ to five-loop order, with $\alpha_s$ given in Eq.(14).}}\label{Fig:spectral}
\end{center}
\end{figure} 
The two parameterizations of the spectral function are clearly very similar, except for the incorrect threshold behaviour of the model of \cite{KM}. Due to the quenching produced by the integration kernels they eventually lead to the same final result for the strange-quark mass. Above $s \simeq 3.5 \; {\mbox{GeV}}^2$ the spectral functions are well accounted for by PQCD, thus suggesting a Cauchy radius $s_0$ in this region. \\

Regarding the OPE for $\psi_5(s)$, the dimension $d=4$ quartic quark mass correction turns out to be negligible, so that there is no need to give up normal ordering in condensates in order to avoid logarithmic quark mass singularities \cite{m4}. Also negligible is the contribution of  the strange-quark condensate. This leaves the dimension $d=2$ quark mass correction, the light-quark condensate \cite{GMOR}: $\langle \bar{q} q \rangle = (- 267 \pm 5 \,{\mbox{MeV}})^3$, the dimension $d=6$ condensate, and the $d=4$ gluon condensate \cite{tau}-\cite{e+e-}:
$\langle \frac{\alpha_s}{\pi} G^2 \rangle = 0.012 \pm 0.012$, where the error combines results from both determinations. The input value  $\bar{m}_s/\bar{m}_{ud} = 27 \, \pm 1$ from \cite{FLAG} has been used.
Regarding the dimension $d=6$ term in the OPE, dominated by the four-quark condensate, there is no experimental data in this channel to determine it, as done e.g. for the vector and axial- vector correlators \cite{tau}-\cite{e+e-}. A  very rough estimate can be obtained by invoking vacuum saturation. This approximation has been shown to break down at next-to-next-to leading order \cite{d=6}, it gives the wrong sign for the ratio of the vector and axial-vector  $d=6$ condensates \cite{tau}-\cite{e+e-}, and it underestimates the vector $d=6$ condensate by a factor of 10 \cite{tau}-\cite{e+e-}. Hence, we assume a $1000\%$ error on the vacuum saturation result, i.e. a factor ten uncertainty. Finally, regarding  the dimension $d=8$ condensate, entering due to the kernel \eqref{EQ:p2}, it
is completely unknown, thus becoming a source of an unknown systematic uncertainty in this case.

There are a number of criteria to determine a suitable value, or range of values for $s_0$. For example, in \cite{CHET_K} the authors find   $s_0=(4.5\pm 0.5)\,\text{GeV}^2$ by demanding stability of the strange quark mass against changes in the Laplace transform parameter. Another criterion used in \cite{KM} is to fix $s_0$ in the region where local duality between the hadronic model and PQCD takes place. From  Fig. \ref{Fig:spectral}  this region is in the range $s_0 = (3.7-4.0)\,\text{GeV}^2$. Beyond this point, the hadronic model Eq. (8)  cannot provide a good description of the spectral function as it  falls off quite rapidly. Another criterion would be to fix $s_0$ by requiring global duality to hold. In this case, integrating both the hadronic spectral function and PQCD with no kernel, i.e. $p(s)=1$, from zero up to $s_0$ one finds agreement for $s_0\approx 4.8\,\text{GeV}^2$. We  consider this ambiguity in fixing $s_0$ as an additional uncertainty, and choose $s_0=(4.2\pm 0.5)\,\text{GeV}^2$, which complies with the latter two criteria.          
\begin{figure}
[ht]
\begin{center}
\includegraphics[height=3.5in, width=4.5in]
{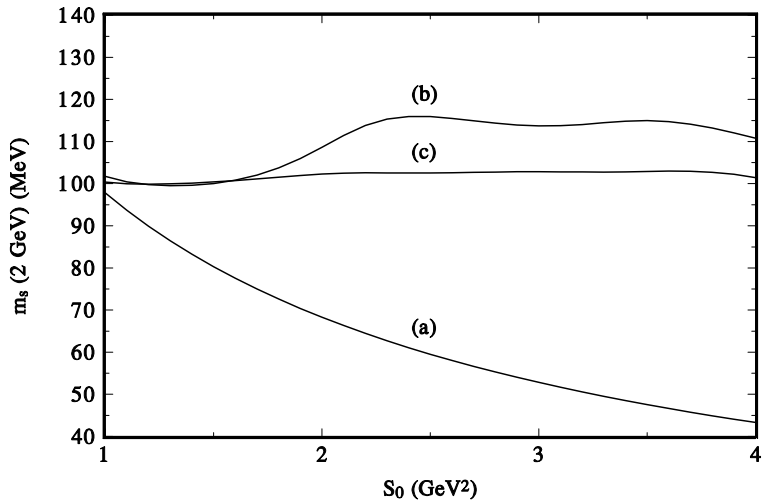}
\caption{\footnotesize{The result for $\bar{m}_s(2 \, {\mbox{GeV}})$ from \cite{DNRS1} using only the kaon pole, curve (a), the full hadronic spectral function, Eq.(9), with no integration kernel ($p(s)=1$), curve (b), and with $p_1(s)$ as in Eq.(5), curve (c).}}
\end{center}
\end{figure}   

Before proceeding to the analysis we wish to briefly comment on the unveiling of systematic uncertainties. Since the quark mass must be independent of the value of the Cauchy radius, $|s_0|$, (provided it is large enough for QCD to be valid), it has been traditional to seek  as much stability as possible in this mass against changes in $|s_0|$. Figure 2 shows results for $\bar{m}_s(2 \, {\mbox{GeV}})$ obtained in \cite{DNRS1} using only the kaon pole in $\psi_5(s)^{HAD}$, curve (a), and the full hadronic spectral function, Eq.(9), but with no integration kernel in Eq.(4), i.e. $p_i(s) = 1$, curve (b). Since this result is reasonably stable for $s_0 \simeq 2.2 - 4.0 \; {\mbox{GeV}}^2$, one would have concluded that $\bar{m}_s \simeq 110 - 118 \; {\mbox{MeV}}$, with a certain error due to other sources. However, using the integration kernel $p_1(s)$, Eq.(5), thus quenching the contribution of the resonance sector, leads to the considerably lower result $\bar{m}_s \simeq 100\; {\mbox{MeV}}$, thus unveiling a systematic uncertainty from the hadronic sector, and acting in only one direction. 
\begin{figure}
[ht]
\begin{center}
\includegraphics[height=2.5in, width=3.5in]
{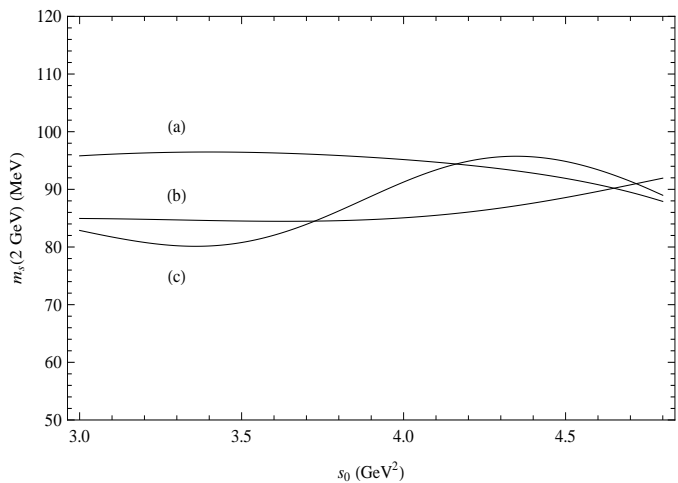}
\caption{\footnotesize{The results for $\bar{m}_s(2 \, {\mbox{GeV}})$ as a function of $s_0$ using the integration kernel $p_3(s)$, Eq. \eqref{EQ:p3}, curve (a), $p_2(s)$, Eq. \eqref{EQ:p2}, curve (b), and $p_1(s)$, Eq. \eqref{EQ:p1}, curve (c).}}
\end{center}
\end{figure}
A separate systematic uncertainty, now in the QCD sector, is suggested by the poor convergence of $\psi_5^{"}(s)$, Eq.(3). The perturbative QCD expansion of $\delta_5^{PQCD}$, Eq.(12), with the integration kernel $p_1(s)$, Eq.\eqref{EQ:p1}, and for $s_0 = 4.2\, {\mbox{GeV}}^{2}$ (with $\mu=\sqrt{s_0}$) is given by
\begin{equation}
\delta_5^{PQCD}= 0.23 \, {\mbox{GeV}}^{8} \left[1 + 2.2 \,\alpha_{s} + 6.7\,\alpha_{s}^2 + 19.5 \,\alpha_{s}^3 + 56.5\, \alpha_{s}^4\right]\;,
\end{equation} 
which after replacing a typical value of $\alpha_s$ leads to all terms beyond the leading order to be roughly the same, e.g. for $\alpha_s = 0.3$ the result is
\begin{equation}
\delta_5^{PQCD}= 0.23 \, {\mbox{GeV}}^{8} \left[1 + 0.65 + 0.60 + 0.53 + 0.46\right]\;,
\end{equation}
which is hardly (if at all) convergent. In fact, judging from the first five terms, this expansion
is worse behaved than the non-convergent harmonic series.
Since the quark mass, Eq.(13), depends on $(\delta_5^{QCD})^{-1/2}$, we find after expanding the latter
\begin{equation}
(\delta_5^{PQCD})^{-1/2} = 2.08\, {\mbox{GeV}}^{-4} \left[1 - 1.10 \,\alpha_{s} - 1.52 \,\alpha_{s}^2 - 2.08 \,\alpha_{s}^3 -  3.21 \,\alpha_{s}^4\right]\;,
\end{equation} 
which exhibits a considerably improved convergence, e.g. for $\alpha_s = 0.3$ this expansion becomes
\begin{equation}
(\delta_5^{PQCD})^{-1/2} = 2.08\, {\mbox{GeV}}^{-5}  \left[1 - 0.33 - 0.14 - 0.06 - 0.03\right]\;.
\end{equation} 
We shall then use Eq.(17) in the sum rule Eq.(13). Figure 3 shows the result for the strange-quark mass at $2\, {\mbox{GeV}}$ using the three kernels $p_1(s), p_2(s)$ and $p_3(s)$, Eqs. (5)-(7). Given the very good stability achieved by the kernel $p_2(s)$, Eq. (6), one would be tempted to prefer it over the other two. However, this result hides an unknown, and potentially very large systematic uncertainty due to the contribution of the unknown dimension $d=8$ condensate. As discussed in connection with Fig. 2, relying exclusively on stability can introduce a sizable systematic uncertainty. For this reason we do not consider this kernel in the sequel, and concentrate on $p_1(s)$ and $p_3(s)$, Eqs. (5) and (7). 
A complete breakdown of the uncertainties for $\bar{m}_s(2 \;{\mbox{GeV}})$, calculated using these two kernels is given in Table 1. While both results are consistent with each other, we choose the one with the lowest uncertainty,  i.e. with the kernel $p_3(s)$, eq. \eqref{EQ:p3}. Not included in the table is the uncertainty due to the value of the hadronic parameter $\lambda$, which controls the relative weight of the second resonance with respect to the first one (see Eq.(9)). Changing the value $\lambda = 1$ by $\pm \, 50\%$ does not change the total error $\Delta_T$, except for the kernel $p_3(s)$,  adding $1 \;{\mbox{MeV}}$ to $\Delta_T$ and resulting in
\begin{equation}\label{EQ:result}
\bar{m}_s( 2{\mbox{GeV}}) = 94 \, \pm 9 \,{\mbox{MeV}} \,.
\end{equation}
Concerning the free parameter $a$ in the kernel $p_3(s)$, Eq. (7), it impacts the uncertainty $\Delta{ s_0}$. For the range $a = 2.6 - 3.3 \; {\mbox{GeV}}^2$ the strange-quark mass changes in the interval
$m_s (2\; {\mbox{GeV}}) = 88 - 100\,\text{MeV}$, in agreement within errors with the value in Eq.(19).
This result should be confronted with that obtained without expanding as in Eq.(17). Using the kernel $p_1(s)$, Eq.(5), and $s_0= 4.2 \; {\mbox{GeV}}^2$ gives
\begin{eqnarray}\label{EQ:resultold}
\bar{m}_s( 2{\;\mbox{GeV}}) &=& [117 \pm 1.6_{\Delta \alpha_s} \pm 1.6_{\Delta_{ \langle G^2\rangle}} \pm 6_{\Delta_{ d=6}} \pm 18_{\Delta_{ s_0}}\pm 5_{\Delta_{ \text{HAD}}}\pm 9_{\Delta_{\text{tr}}}] {\mbox{MeV}}\nonumber \\ [.3cm]
&=&
117 \, \pm 22 \,{\mbox{MeV}} \,,
\end{eqnarray}
where the uncertainties have the same meaning as in Table 1. This result differs somewhat from that of \cite{DNRS1}, Eq.(1), obtained with the same kernel, because of different values used here for the various input parameters. Due to the huge uncertainty from the choice of $s_0$ this result agrees within errors with that given in Table 1 using the same kernel.
Using instead the kernel $p_3(s)$, Eq.(7), and still not expanding as in Eq.(17), gives
\begin{eqnarray}\label{EQ:resultnonexp}
\bar{m}_s( 2{\;\mbox{GeV}}) &=& [125 \pm 1.8_{\Delta \alpha_s} \pm 1.5_{\Delta_{ \langle G^2\rangle}} \pm 4_{\Delta_{ d=6}} \pm 5_{\Delta_{ s_0}}\pm 8_{\Delta_{ \text{HAD}}}\pm 12_{\Delta_{\text{tr}}}]\,{\mbox{MeV}} \nonumber \\ [.3cm]
&=&
125 \, \pm 16 \,{\mbox{MeV}} \,,
\end{eqnarray}
which is inconsistent with Eq.(19). It is conceivable that the difference originates from an underestimation of the truncation error, thus leading to a large systematic uncertainty. For a poorly convergent series, such as Eq.(16), it is dangerous to estimate the unknown term from the difference between the last two known ones. This clearly shows the importance of keeping convergence under control, i.e. results change drastically when convergence is notably improved as in Eq.(17).\\  

\begin{table}
\begin{tabular}{cccccccccc}
& \multicolumn{8}{c}{$\;\;\;\;\;\;\;\;\;\;\;\;\;\;\;\;\;\;\;\;$Uncertainties (MeV)}\\
\cline{3-9}
\noalign{\smallskip}
Kernel 	&	$\bar{m}_s(2\,\text{GeV})$	& $\Delta_{\alpha_s}$ 	& $\Delta_{ \langle G^2\rangle}$ 	& 	$\Delta_{ d=6}$ 	& $\Delta_{ s_0}$ 	&		$\Delta_{ \text{HAD}}$ 	&	$\Delta_{\text{tr}}	$	& $\Delta_{\text{T}}$ \\   
& (MeV)&&&&&&&\\
\hline 
\noalign{\smallskip}
$p_1(s)$									&			95		 																		&		3								& 0.3						& 1									&  10 						& 4				& 6				&	13			\\

$p_3(s)$									&			94		 																		&		4								& 0.6						& 1									&  3 						& 6				& 2				&	8			\\      
\hline
\end{tabular}
\caption{\footnotesize{Results using kernels $p_1(s)$ and $p_3(s)$ with $a=2.8\,\text{GeV}^2$, Eqs. (5) and
(7), for  $s_0= \mu^2 =(4.2\pm 0.5)\,\text{GeV}^2$. The uncertainties are due to the
 values of $\alpha_s$ ($\Delta_{\alpha_s}$), the gluon condensate ($\Delta_{\langle G^2\rangle}$), 
 the four-quark condensate ($\Delta_{d=6}$), the range in $s_0$ ($\Delta_{s_0}$), the hadronic model
  ($\Delta_{\text{HAD}}$), and the truncation of PQCD ($\Delta_{\text{tr}}$), combined in quadrature to give the total uncertainty ($\Delta_{\text{T}}$).}}
\end{table}

In order to asses the impact of a higher pinching on the $s_0$ dependence of $m_s$ we consider the kernel

\begin{equation}
p_4(s) = (a - s) (s_0 - s)^2 \;,
\end{equation}
 which leads to

\begin{eqnarray}
\bar{m}_s( 2{\;\mbox{GeV}}) &=& [94 \pm 4_{\Delta \alpha_s} \pm 0.1_{\Delta_{ \langle G^2\rangle}} \pm 1_{\Delta_{ d=6}} \pm 0.3_{\Delta_{ s_0}}\pm 5_{\Delta_{ \text{HAD}}}\pm 5_{\Delta_{\text{tr}}}] \,{\mbox{MeV}}\nonumber \\ [.3cm]
&=&
94 \, \pm 8 \,{\mbox{MeV}} \,.
\end{eqnarray}

This result is in remarkable agreement with those in Table 1 using the kernels $p_1(s)$ and $p_3(s)$. However, this kernel would involve the contribution of vacuum condensates of dimension $d=8$, which have not been taken into account in Eq.(23), as they are mostly unknown. Judging from the size of the uncertainty due to the dimension $d=6$ condensate in Eq.(23), one should  expect this result to be reasonable reliable.

\section{Conclusions}
In this paper we reconsidered a previous strange-quark mass determination \cite{DNRS1} taking into account additional potential systematic uncertainties. These uncertainties arise mostly from the poor convergence of the pseudoscalar correlator in PQCD, and the hadronic parametrization of the spectral function. The former has been uncovered by using an alternative expansion of the PQCD correlator exhibiting a considerably better convergence. The latter has been addressed by considering integration kernels with improved quenching of the resonance sector. Updated values of the various quantities  entering the pseudoscalar correlator have been used, and a generous error given to the dimension $d=6$ condensate. The result is a consistent picture leading to the value in Eq.(19), as the one with the least overall uncertainty. This value is in good agreement with contemporary QCD sum rule determinations to five-loop order in PQCD using the scalar correlator \cite{Other}, and from Laplace transform sum rules for the pseudoscalar correlator \cite{CHET_K}. It is also in agreement with various LQCD results \cite{FLAG}. In comparison with the previous determination in a similar framework, Eq. (1), it must be pointed out that the latter is affected by a systematic error due to the poor convergence of PQCD, as appreciated from Eqs. (15)-(16).  The new result obtained here, Eq.(19), while in agreement within errors with Eq.(1), should be much less affected  by that uncertainty.

\end{document}